\documentclass{article}
\usepackage{graphicx}

\begin{document}
\title{Merging quantum theory into classical physics.
\author{Jacques Moret-Bailly 
\footnote{Laboratoire de physique, Universit\'e de Bourgogne, BP 47870, F-21078 Dijon cedex, France. email : 
Jacques.Moret-Bailly@u-bourgogne.fr}}}
\maketitle

\begin{abstract}
The power of quantum mechanics is not a proof of the worth of its postulates criticised by Ehrenfest, Einstein, 
Schr\"odinger, deBroglie and more recent authors. Rejecting its postulates, the useful remainder may be 
considered as part of a classical theory including Planck's quantization and de Broglie's waves.

The existence of the zero-point electromagnetic field is deduced from classical electrodynamics, while its intensity 
requires Planck's constant. Taking this field into account, classical electromagnetism shows that an electron 
turning around the kernel of an atom may keep its energy.

The zero point field is an ordinary field existing in the dark, which cannot be separated from the total 
electromagnetic field in an excited mode. The total field is in equilibrium with matter that it polarizes temporarily 
and reversibly. This polarisation is exceptionally large enough to allow the energy of an atom reach and cross a 
pass between two minimums of potential, stimulating an emission or an absorption.

A paradox of quantum electrodynamics is explained classically: The excitation of an atom corresponds to the 
emission of a spherical wave which cancels partly an external wave; the nearly plane wave generally used in 
stimulated emission experiments must be decomposed into an efficient spherical wave and a scattered wave, so 
that the plane wave is two times less efficient than the zero point component of the spherical mode.

The photon is not a particle, only a quantum of electromagnetic energy exchanged by a system performing a 
transition between stationary states. The classical electromagnetism does not need the quantum postulate 
\textquotedblleft reduction of the wave packet"; it explains how an absorption (or emission) may be coherent 
whereas the atoms perform quantified transitions; it has no paradox such as EPR.

De Broglie's waves may be the linear part of solitons of a field which may be a high frequency electromagnetic 
field. These solitons may lead to a physical interpretation of the superstrings theory.

A part of the quantum computations of the stationary values of the energy results from natural or assumed 
symmetries; the remainder requires so many arbitrary hypothesis that the computations may be considered as a 
phenomenological component of classical mechanics.
\end{abstract}

\section{Introduction}
Analytical mechanics and geometrical optics lead to the same variational equations. The propagation of particles 
such as the electron, the atoms leads to the observation of interferences. Thus the problem of the wave particle 
duality appears. Quantum mechanics claims that it solves this problem; it allows the physicist consider arbitrarily 
that he studies a wave or a particle; but an importance of the observer in the study of a physical problem is often 
considered shocking; the worst is that, leading to the same variational properties only in a first approximation, the 
wave equations are incompatible with the propagation of the particles. Consequently quantum theory introduces 
a postulate, \textquotedblleft the reduction of the wave packet" to shift arbitrarily from a wave solution to an 
other.

The two main theories of physics in the twentieth century, relativity and quantum mechanics were strongly 
discussed; but, while the disputes about relativity decreased quickly, quantum mechanics remained for some 
people so strange or absurd that it appeared necessary for its defenders to imagine a lot of \textquotedblleft 
definitive tests" such as fourth order interference experiments. The tests work, but they do not bring the expected 
proof.

Einstein, Schr\"odinger and others imagined experiments for which the quantum interpretation may be absurd, 
showing \textquotedblleft paradoxes of quantum mechanics". Lots of discussions followed, without any clear 
conclusion, as their authors use obscure concepts of quantum mechanics : for instance they attach particles of 
light named photons to modes of the electromagnetic field which are not well defined. A similar obscurity led a lot 
of specialists of quantum theory to deny the possibility that the lasers work\cite{Townes}.

\medskip
This paper is not a review of the criticisms of quantum mechanics: its aim is showing that the classical theory is 
able to explain the effects whose strangeness was at the source of quantum theory, and that the useful part of 
quantum theory may be considered as phenomenological, thus included into the classical theory.

\medskip
Section 2 shows that classical electrodynamics (CED) is as able to explain all experiments of optics than quantum 
electrodynamics (QED), with the advantage of replacing postulates by demonstrations. It uses two properties of 
the classical fields which are trivial, but often neglected: the difficult absorption of a wave well known in 
acoustics, and the interference of the electromagnetic fields which disconnects them from the energy.

Subsection 2.1 is a simple recall of the general definition of the sets of modes of an electromagnetic field, showing 
the problems bound to the definition of the photon in QED.

Subsection 2.2 studies the classical interaction of a wave with a single atom.

Subsection 2.3 introduces classically the zero point electromagnetic field with which the consequences of CED 
appear equivalent to those of QED \cite{Welton,Boyer,Marshall,Marshall1,Marshall2}.

Subsection 2.4 shows that the experiments of optics which try to demonstrate that CED is wrong are founded on 
too bad approximations.

Subsection 2.5 shows how the quantum postulate \textquotedblleft reduction of the wave packet" leads to a 
failure in the comparison of spontaneous and stimulated emissions.

Section 2.6 describes the classical interpretation of the emission and absorption of a quantum of electromagnetic 
energy by a set of particles.

\medskip
Section 3 introduces a tool which enables to solve classically, really the wave-particle duality: an example of 3-
dimensional, static soliton.

Subsection 3.1 recalls experimental and theoretical results of nonlinear optics, more precisely the properties of the 
filaments of light commonly observed in the propagation of powerful laser pulses through matter.

Subsection 3.2 shows that light filaments can keep several configurations if a magnetic nonlinearity is introduced; 
in particular, the filaments may curve and splice: most of their energy is confined in a limited region of the space, 
possibly static. It is (3+0)D solitons.

\medskip
Section 4 proposes a tentative model of particles by a classical explanation of de Broglie's \textquotedblleft 
double solution" through the hypothesis that the solutions are the quasi-linear and nonlinear parts of the field 
making a soliton.

Subsection 4.1 sets that at high energies and frequencies, the vacuum becomes non-linear, so that the previously 
introduced solitons may appear in the vacuum; the field of these solitons is split into the two fields of de Broglie's 
\textquotedblleft double solution"; the superstrings and such light filaments may be identified.

The existence of eigenvalues of the energy, that is the existence of relative minimums of a potential function is not 
a problem in classical mechanics, while the computation of these energies may be difficult. Subsection 4.2 
analyses the main application of quantum theory, the computation of the eigenvalues of the energy, to consider 
these computations as phenomenological, thus, possibly, classical.

\section{Comparison of classical and quantum electrodynamics}
Quantum electrodynamics is set by an identification of the modes of the electromagnetic field with quantum 
oscillators. Various systems of mode are used: plane monochromatic waves in a cube, infinite plane 
monochromatic waves \dots Attaching to a particle \textquotedblleft photon" the electromagnetic wave which 
explains the propagation (and seems used as Schr\"odinger's wave function) is not permanently possible, so that 
the postulate \textquotedblleft reduction of the wave packet" is necessary. Does it solves all problems? Is 
classical electrodynamics able to explain all observations ? It is necessary to set precisely what are the optical 
modes and how light interacts with a molecule \footnote{\textquotedblleft Atom", \textquotedblleft molecule" are 
used for any small set of atoms able to emit or absorb light in a gas, a liquid, a crystal or an other solid.}.

\subsection{Modes of the electromagnetic waves.}
In the vacuum, the electric component of the electromagnetic field obeys the second order linear vector equation:
\begin{equation}
(\nabla^2 -\epsilon_0\mu_0\partial^2/\partial t^2){\bf E}=0
\end{equation}
For chosen conditions at the limits, the solutions of this equation build a vector space. Suppose that the energy 
at a given instant, in the (physically meaningless) absence of any other mode can be computed; two modes are 
orthogonal if the energy of the sum of the fields is the sum of the energies of the fields of each mode. A complete 
set of orthogonal solutions defines a rectangular frame. Such a set is named a \textquotedblleft set of modes"; 
any solution of the equation, named \textquotedblleft mode" is evidently a linear combination of the elements of 
the set.

An infinity of conditions at the limits, and of sets of modes is available. A Fourier representation is often chosen; 
it may be a Fourier transform for all variables, or, as Planck did, Fourier series for space variables, in a cube. Using 
developments in wavelets to define the set provides a more physical frame.

An inconvenient of the Fourier transform is that a normalisation of the energy of a sinusoidal mode at Planck's 
quantum $h\nu$ is impossible. The use of a box and a limitation of the time at a period is an acceptable method to 
avoid the more rigorous treatment by wavelets. An other trick is computing the flux of the vector of Poynting 
through a wave surface.

However, an experiment is always limited in the time, so that a monochromatic wave and consequently the law 
$W=h\nu$ is a limit case. The most general, more meaningful law 
\begin{equation}
\int \frac{w(\nu)d\nu}{\nu}=h,\hskip 5mm {\rm with \hskip 1mm an \hskip 1mm exchanged \hskip 1mm 
energy}\hskip 5mm W=\int w(\nu)d\nu,.\label{hnu}
\end{equation}
reduces to the ordinary law for a monochromatic wave.

\medskip
The strict theory of modes works in a transparent medium, while optics needs sources, that is charges which 
destroy the propagation equation (1). However, the sources are generally small, so that equation (1) works almost 
everywhere. The following extension must be used carefully:

The field ${\bf E'}$ emitted by a source is transformed by a charge and time inversion into a completely absorbed 
field ${\bf E"}$. Summing both problems the source disappears and the field ${\bf E'+E"}$ verifies Maxwell's 
equations, it makes a mode. If ${\bf E"}$ disappeared before an experiment, it is possible to replace it by the mode 
to make computations; thus ${\bf E'}$ defines the \textquotedblleft emission mode" of the source. An absorption 
mode is defined similarly from ${\bf E"}$.

\subsection{Elementary light-matter interaction in classical optics.}
In classical optics, a dipole radiates a well defined electromagnetic wave named \textquotedblleft spherical" 
although it is not invariant by all rotations around all axis crossing the dipole \footnote{Light-matter interactions 
can use other multipoles: reading \textquotedblleft dipole", understand \textquotedblleft convenient multipole" 
and reading \textquotedblleft spherical wave", understand \textquotedblleft wave radiated by the convenient 
multipole".}. This field is not quantified.

Generally, an emitted field is partly cancelled by an external field; a source of field is a source or a receiver of 
energy in function of the external fields. The difference between absorption and emission results only on the 
interference of the emitted spherical field with external fields; a limiting case is no exchange of energy, an 
equilibrium between the atoms and the external field which explains that, in Bohr's model, the electrons of an atom 
do not fall to the kernel.

\medskip
 It is fundamental to distinguish well two types of interactions:

i) The polarisation is a non-quantified, temporary and reversible absorption of energy, in which the 
potential energy of the atom remains close to a minimum (eigenvalue) of the potential. The polarization 
by light of dense enough a matter may be observed using, for instance, the dynamical Kerr effect, 
without an interaction with the exciting electromagnetic field.

ii) If a polarisation is large enough, the potential energy may reach and cross a pass to an other minimum of 
potential, higher or lower than the initial one. Thus a transition occurs. If the initial and final states are 
stationary, that is if they correspond to potentials close to the minimums, the quantification relation 
(\ref{hnu}) applies.

\medskip
Quantum electrodynamics (QED) identifies the modes of an arbitrarily chosen set with harmonic oscillators. A set 
of modes chosen to contain the mode of emission of a source does not contain generally the absorption mode of 
a receiver, but the two modes must be not orthogonal. The postulate of QED \textquotedblleft reduction of the 
wave packet" is used to transform arbitrarily the emission mode into the absorption one\dots. Classical physics, 
does not use such a postulate, the required demonstration is given in subsection 2.6.
\subsection{The classical zero point field.}

Studying the emission of an electromagnetic wave, it is often, implicitly supposed that it is alone in the universe, 
so that it appears that energy is always emitted, and energy seems simply bound to the field.

A full absorption of a wave requires a full cancellation of this wave by the field emitted by receivers. The 
cancellation of a noise by the emission of a convenient field is a problem well known in acoustics, a difficult 
problem. In electromagnetism, it is easy to imagine the cancellation of a plane wave by other plane waves, but the 
cancellation of the wave emitted by a small source is very difficult: it requires the building of a wave identical to 
the wave radiated by the source, except for a $\pi$ phase.

The electromagnetic wave emitted by a nearly punctual source as an atom is generally developed using spherical 
harmonics. The experiments show that, with a good approximation, a single spherical harmonic is sufficient; 
consider, for instance, dipolar emissions.

The field radiated by a dipolar source at a point $O$ is singular at $O$, while the field radiated by other nearly 
punctual sources is regular at this point, so that the absorption requires almost an infinity of absorbers, thus an 
large time. Thus, it remains a field named \textquotedblleft zero point field".

Consider an atom and the spherical harmonic monochromatic mode of emission or absorption of a photon by this 
atom. Suppose that the atom is in a cold blackbody, so that it may emit, but not get photons. If the energy in the 
mode is between 0 and $h\nu$, nothing happens. If the energy is between $h\nu$ and $2h\nu$, the atom may 
radiate a photon, so that in both cases, the final energy reaches a value between 0 and $h\nu$. Supposing that 
the initial value of the energy is arbitrary, the mean final value is $h\nu/2$, the value computed by Planck 
\cite{Planck} and Nernst \cite{Nernst}. 

The previous explanation is not sufficient because it is not sure that the modes have arbitrary initial modes; 
verifying the value $h\nu/2$ will correct simultaneously the first Planck's law into the second. At a high 
temperature the first Planck's law sets the energy of a mode:
\begin{equation}
\frac{h\nu}{exp(h\nu/kT)-1}\approx\frac{h\nu}{1+h\nu/kT+(1/2)(h\nu/kT)^2-1}\approx kT-\frac{h\nu}{2}
\end{equation}
Thermodynamics shows that a degree of liberty, here the mode must have the energy $kT$ for $T$ large; the 
necessary addition of $h\nu/2$ transforms the first Planck's law into the second, adding the zero point energy of 
the mode\footnote{A philosopher may search the origin of Planck's constant:

- either the Universe sets the zero point field, and most atoms get a dynamical structure in equilibrium with this 
field;

- or $h$ results from infinitesimal properties of the elementary particles which fill the Universe of 
electromagnetic field up to the equilibrium.}.

 The word \textquotedblleft photon" is the name of a particle in QED. When this word is used in CED, its meaning 
is \textquotedblleft quantity of energy $h\nu$" , not particle, so that there is no problem of \textquotedblleft wave-
particle duality" for the electromagnetic waves. The photon of QED is defined in a particular mode; to avoid a 
splitting of the particle by a splitting of the wave into other modes, its mode may be arbitrarily changed.

It is fundamental to remark that the zero point field, although it is residual, is an absolutely ordinary field directly 
detected by the Casimir attraction forces \cite{Casimir}. But the zero point field exists only in the dark; after its 
amplification by a source, it does not exist anymore, only the total field is meaningful. However studying the 
energy provided by a spontaneous source (similarly an absorbed energy), we may subtract the energy of the zero 
point field ${\bf Z}$ {\it existing before} the amplification, from the energy of the (total) field ${\bf E}$ {\it 
existing after} the amplification, obtaining an energy proportional to $\bf E^2-\bf Z^2 = {\bf (E-Z)(E+Z)}\approx 
2{\bf Z(E-Z)}$ for a weak source; this formula introduces ${\bf E-Z}$ named \textquotedblleft usual field" or 
\textquotedblleft conventional field" although it is not a physical field, its components ${\bf E}$ and ${\bf Z}$ 
existing in different points. It is improper but comfortable to write that the total field is split into a zero point field 
and the conventional field.

The equivalence between the quantum theory and a classical theory which does not neglect the zero point field is 
established in particular by Marshall and Santos \cite{Marshall}. These authors use the name \textquotedblleft 
stochastic field" rather than \textquotedblleft zero point field" but this does not seem good because this field 
exists only in the dark, and the meaningful total field is shaped by the sources. 

\subsection{The zero point field and the detection of low level light.}
Considering absorptions in this section, it will be necessary to compare the total, initial field to the remaining zero 
point field, so that, by subtraction it {\it seems} that the total field is split into a zero point field and a remainder, 
the conventional field. Thus, for a mode, set $\bf E$ the total electromagnetic field (before absorption), $\bf Z$ the 
zero point field (after absorption) and $\bf F$ their difference, the conventional field .

The available energy in a mode is the volume integral of $(\epsilon_0{\bf E}^2+\mu_0{\bf H}^2)/2$ or 
$\epsilon_0{\bf E}^2$ computed at a given instant\footnote{The parameters of the vacuum $\epsilon_0$ and 
$\mu_0$ are used because the polarisation of the atoms is introduced explicitly.}. Suppose that this energy is 
absorbed by cooled atoms until the equilibrium is reached. The mode loses an energy proportional to 
$\epsilon_0E^2-\epsilon_0 Z^2$. This result is usually approximated neglecting the zero point field, but if the 
energy is low $2\epsilon_0 {\bf (E-Z).Z}$, that is the interference of the conventional field $\bf F=(E-Z)$ with the 
zero point field, must not be neglected.

\begin{figure}
\begin{center}
\includegraphics[height=3 cm]{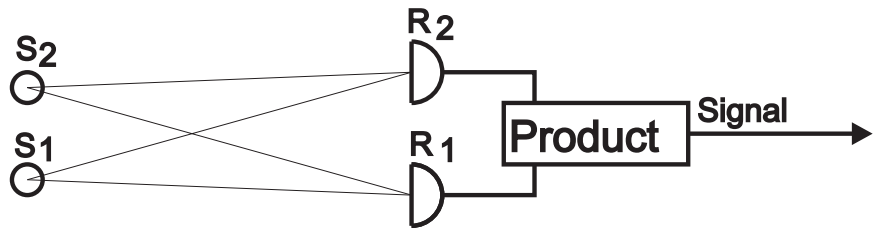}
\end{center}
\caption{Fourth order interferences.}
\end{figure}

Neglecting the zero point field in the computation of fourth order interference experiments, many experimenters 
thought that the classical theory failed, hoping to demonstrate a superiority of QED \cite{Gosh,Ou1,Ou2,Kiess}. 
In the simplest version of the experiment\cite{Mandel}, two small monochromatic, incoherent sources $S_1$ and 
$S_2$ light equally two small receivers $R_1$ and $R_2$ (figure 1). The signal is the product of the detected 
intensities $I_1$ and $I_2$ . Set $\lambda$ the wavelength, $\phi$ the instantaneous, quickly variable difference 
of phase of the two sources, ${\bf Z}$ the amplitude of the zero point field in the mode absorbed by a receiver 
and ${\bf F}$ the conventional amplitude sent by the sources at $R_1$ or $R_2$; the product of the detected 
intensities is:
$$I_1I_2 =\Bigl[{\bf Z+F} \cos\Bigl(\frac{\pi(S_1R_1-S_2R_1)}{\lambda}+\frac{\phi}{2}\Bigr)\Bigr]^2\Bigl[{\bf 
Z+F}\cos\Bigl(\frac{\pi(S_1R_2-S_2R_2)}{\lambda}+\frac{\phi}{2})\Bigr)\Bigr]^2
$$
Supposing that the sources are weak, ${\bf F}$ is small, compared to ${\bf Z}$; as the mean value of the cosines 
containing the fast variating phase $\phi$ is zero:
$$
I_1I_2 \approx {\bf Z}^4+4({\bf ZF})^2 \Bigl[\cos\Bigl(\frac{\pi(S_1R_1-
S_2R_1)}{\lambda}+\frac{\phi}{2}\Bigr)\cos\Bigl(\frac{\pi(S_1R_2-
S_2R_2)}{\lambda}+\frac{\phi}{2})\Bigr)\Bigr]
$$
${\bf Z}^4$ is not taken into account because it exists if the source is switched off \footnote{However it is a 
source of noise.}. In the average $({\bf ZF})^4$ is a positive constant. Setting $\Delta=S_1R_1-S_2R_1-
S_1R_2+S_2R_2$, the signal is proportional to
$$
\cos\frac{\pi\Delta}{\lambda}+\cos\Bigl(\frac{\pi(S_1R_1-S_2R_1+S_1R_2-S_2R_2)}{\lambda}+\phi\Bigr).
$$
As the difference of the phases of the incoherent sources $\phi$ changes quickly, the second cosine is averaged 
at zero, so that the signal is proportional to $\cos(\pi\Delta/\lambda)$.

The visibility of the fringes tends toward 1 as the intensity tends to zero: QED is not necessary to explain the 
experimental result.

\medskip

\begin{figure}
\begin{center}
\includegraphics[height=6 cm]{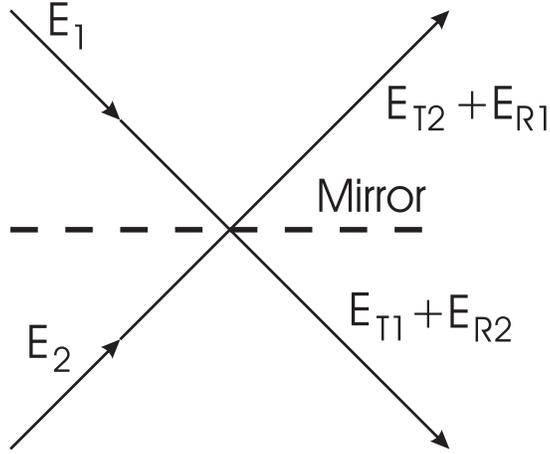}
\end{center}
\caption{Interference of beams using semi-reflecting mirrors.}
\end{figure}

An other type of experiments trying to show the superiority of QED uses two identical sources and a half 
reflecting beam splitter which mixes the mode emitted by a source and reflected, with the mode emitted by the 
other source and transmitted (figure 2). Set $E_1$, $E_2$ the fields arriving on the splitter, $ E_{T1}, E_{T2}$ the 
emerging transmitted fields and $E_{R1}, E_{R2}$ the reflected fields. The conservation of the fields and of the 
energy on the beam splitter writes for a source:
$$
E_1=E_{R1}+E_{T1} \hskip 5mm E_1^2=E_{R1}^2+E_{T1}^2
$$

These conditions require a difference of phase of $\pi/2$ between $E_{R1}$ and $E_{T1}$ on the surface of the 
beam splitter, that is $\phi_{R1}-\phi_{T1} = \pi/2 = \phi_{R2}-\phi_{T2} $\footnote{The beam splitter has a 
thickness; its \textquotedblleft surface" being defined arbitrarily, a change of the definition may change the 
difference of phases by $3\pi/2$ rather than $\pi/2$, for both beams; a real beam splitter has losses; taking this 
into account does not change the result.}.

A detection is obtained for a large emerging field; this requires, in particular, almost the same phase  $\phi_{T1} 
\approx \phi_{R2}$ for $E_{T1}$ and $E_{R2}$; then  $\phi_{R1}-\phi_{T2} \approx \pi$, the phases for$ 
E_{R1}$ and $E_{T2}$ are opposite, so that the fields subtract at the other output. This result is called 
\textquotedblleft anti-bunching" if one of the sources emits only a zero point field \cite{Carmichael,Kimble1, 
Kimble2} and \textquotedblleft coalescence of photons" if both sources are switched on \cite{Hong,Santori}.
\subsection{Spontaneous emission and absorption: Einstein's coefficients.}
The introduction of the zero point field into Einstein's theory of the stimulated emission \cite{Einstein} leads to 
suppose that the spontaneous emission is an amplification; however many authors remarked that the 
amplification of the zero point field should be two times larger than the amplification of an other field 
\cite{Weisskopf,Ginzburg,Milonni,Milonni2}. These authors explain this discrepancy by a  mostly classical, 
computation of the \textquotedblleft radiation reaction", interaction of a dipole with the field it emits. They try to 
split the total exciting field into two components (zero point and conventional fields), a very arbitrary splitting for 
an amplification in which the zero point field does not appear; then they try to explain different properties for 
these fields; the following classical computations shows their {\it ad hoc} explanations are false.

\medskip
Show that classical optics solves easily this problem, providing a physical splitting of the exciting field.

Study the flux of the vector of Poynting of electromagnetic fields going to the dipole (input fields), through a 
sphere whose centre $O$ contains an excited, oscillating dipole oriented along $z'Oz$. The electric components 
along $x'Ox$ or $y'Oy$ of an external electromagnetic field do not interact with the dipole which is parallel to 
$z'Oz$, their scalar product with the field $\bf E$ radiated by the dipole, is zero. Thus we may consider only 
external fields polarised in the direction of $z'Oz$ and having the frequency of the dipole.

At a point of the sphere, the instantaneous magnetic field, $\bf H$ may be split into its components $\bf {H_x, 
H_y}$, to give with the electric field $\bf E$ the two components $\bf P_y$ and $\bf P_{x'}$ of the vector of 
Poynting.

If the input field is in the \textquotedblleft spherical" mode which converges to the dipole, it is invariant by the 
rotations around $z'Oz$, so that the total flux of $\bf P_{x'}$ and $\bf P_y$ through the sphere are equal and add.

If the input field is produced by a plane wave, $\bf H_{x}$ for instance is identical to zero, so that the variation of 
flux due to the external field is only produced by $\bf P_{x'}$ and is two times lower than in the previous case.

Thus, the dipole absorbs exactly the same energy from an $h\nu/2$ input energy in its spherical, mode than from 
an $h\nu$ input energy of a plane wave. The origin of this paradox is that the plane wave plus the dipole does not 
make a solution of Maxwell's equations; the diffracted output wave corresponding to the radiation reaction must 
be taken into account, it reduces the field at the dipole by the required factor $\sqrt{2}$.

\medskip

A dipole emits or absorbs \textquotedblleft spherical" waves, its two proper modes, without any diffraction. It 
interacts with other modes through their projections on its proper modes: The previous result may be found by a 
decomposition of the plane wave into a spherical mode andorthogonal  diffracted modes.

\medskip
To emit energy, an atom must be polarised up to a threshold; the exciting, spherical mode provides an energy 
$h\nu/2$, plus an energy due eventually to an addition of the electric field of the component of an other mode on 
the proper absorption mode. If the interferences are negligible, so that the energies add, this electric field provides 
half of the energy in its mode: it is not the zero point field which is very efficient, but the external field which is 
not.
 
The quantum \textquotedblleft reduction of the wave packet" hides the modes and leads to a paradox.

\subsection{Mechanism of emission and absorption of a photon.}
Einstein's interpretation of the photoelectric effect and the whole spectroscopy have led to introduce the photon 
as a particle. However these experiments do not prove a quantification of the light: they prove only that if a 
transition occurs {\it between two stationary states}, the exchange of energy between matter and light is 
quantified; in classical physics, this quantization is imposed by the matter.

A stronger argument in favour of the photon is the excitation of atoms by very weak beams which, neglecting the 
fluctuations of the total field, provide the necessary energy slowly and on a large surface. To preserve the 
computation of the electromagnetic field which plays the role of the wave function of the photon, QED uses a 
magic stick, the \textquotedblleft reduction of the wave packet", which transforms a wave emitted spherically 
around a source into a spherical wave converging to an other point. Classical physics needs a demonstration.

\medskip
The following explanations are only a refinement of the general radiation conditions which we recall again here:

{\bf Radiating an electromagnetic field does not mean radiating energy; while a dipole excited by a plane wave 
radiates a \textquotedblleft spherical field", the interferences of both fields lead to radiation of energy mostly in 
the direction of propagation of the exciting wave}. This directivity of the propagation of energy is increased by a 
collective behaviour of dipoles, and, at a low level, by the replacement of the conventional field by the total field.

\medskip
The threshold between two minimums of the potential function of an atom may be very low, so that the 
fluctuations of the zero point field alone excite it: the good photoelectric cells detect these fluctuations even at 
low temperature, in the dark, as noise.
A low increase of the field is sufficient to increase the number of detected fluctuations, providing a mixture of 
signal and noise.

The most common exchanges of energy between light and matter occur by polarization: the propagation of light in 
a transparent medium polarizes it; this polarization is the source of coherent effects, as the refraction and the 
Bragg scattering; this polarisation is directly observed using a probe beam, leading to a non-destructive 
observation of light.

How can a medium be transparent while its polarization requires energy? There is an equilibrium between the 
density of energy of the light and the intensity of the polarization, so that energy is absorbed at the beginning of 
a light pulse, when the electromagnetic energy increases, and it is returned coherently at the end of the pulse. The 
polarised atoms are able to amplify light in modes which are not orthogonal to their emission mode, generally a 
\textquotedblleft spherical" mode of dipolar emission. The observed coherence has two origins: mainly all atoms 
of the medium on a wave surface keep the same phase, reconstituting the exciting wave surfaces by the 
Huyghens' process. If an atom loses its phase, the out of phase component of its radiated field does not add to a 
high field, so that it takes few energy. However these variations of phases, the discontinuity of the media which 
make Huyghens' process approximate, various fluctuations, lead to low losses, the Rayleigh incoherent diffusion.

In classical mechanics, the eigenvalues of energy are relative minimums of energy, and it is assumed that an 
excitation corresponding to a transition resonates at Planck's frequency. This excitation, which corresponds to a 
polarization of the medium is a non linear function of the exciting field, so that it may become catastrophic, 
producing a transition. In an absorbing medium, the catastrophe is allowed to appear for an exceptionally high 
fluctuation of the total field. The spherical wave emitted by the atom which performs a transition absorbing a 
photon corresponds to the emission of a negative energy. The emitted wave is widely cancelled by the field 
which excited the atom, the remainder is spread locally, it does not scatter much energy: the absorption is 
coherent. Remark that, except for a sign, a similar process occurs in the coherent amplification by a laser medium.

While they transfer energy to the atom which will be excited, the atoms of the medium transfer a part of the 
momentum they got while they were excited. The transfer may be large in a low pressure gas, so that the recoil 
and the recoil energy are nearly the same than for an interaction of a single atom, or spread, low, as in the 
M\"ossbauer effect.

Thus, the coherence of emission and absorption results mainly of fluctuations of the zero point field, less of local 
diffusions of the energy. It is the classical \textquotedblleft reduction of the wave packet".
\subsection{Comparison of quantum and classical electrodynamics}
The previous sections show that the experiments set to demonstrate a superiority of QED over CED are well 
explained by both theories. However the basic rules needed to set CED are the macroscopic rules set in the 
nineteenth century, while QED requires additional postulates.

The fundamental difference between both theories is the lack of quantization of the light in CED: while the number 
of photons in a pulse of light has any value in CED, it is an integer in QED; while the emission of a photon is 
followed by its absorption in QED, in CED the emission or absorption of a photon changes the total 
electromagnetic field which includes the zero point field, so that the probability for an emission or absorption of a 
photon by a large fluctuation of the field is changed; in the average, the emission of a photon leads to the 
absorption of a photon, but the spread and the fluctuations of the field allow the absorption of zero, two or more 
photons. A study of the interaction between an electromagnetic field confined in a box and a set of two levels 
atoms having nonlinear dipoles shows that the atoms perform transitions to keep the mean density of energy in 
the field \cite{Monnot}. As there is not a one to one correspondence between the emissions and absorptions, the 
Einstein, Podolsky and Rosen experiment is meaningless in CED.

\medskip
Quantum cryptography, founded on the quantization of the light or EPR experiment cannot work in CED: thus 
testing quantum cryptography would be a test allowing a valuable choice.

A lot of experiments of quantum cryptography have been proposed, many experimenters claimed that it works. 
But while the experiments showed that it is possible to send a code using low level light, no experiment showed 
the impregnability of the transmission using \textquotedblleft blind" operators. Two experiments could be used :

- A first argument for quantum cryptography is that any measure perturbs the measured object, here a pulse of 
light. But the quantum theory of measure seems to work only if transitions occur. Suppose that a spy 
measures the polarisations of the fibre in two or more directions to detect the two components of the field. Is 
the legal receiver able to detect the presence of the spy ?

- If the Copenhagen theory of the measure fails, the spy cannot be detected, but an EPR experiment would 
change the characteristics of the photon measured by the spy at its detection by the legal receiver or at the 
measure of its twin by the sender. If the spy amplifies, then splits the signal, he can use exactly the same 
detecting apparatus than the legal receiver; is it any difference between them?
\section{Some properties of nonlinear waves: The (3+0)D solitons.}
In the first part of the paper, the classical problem was simple considering that the photon is not a particle. 
Consider now massive particles and their de Broglie's waves: the problem of the wave-particle duality appears. De 
Broglie started its resolution, introducing the \textquotedblleft double solution" \cite{deBroglie}, but his attempts 
to bind the solutions by a nonlinear interaction failed. Experiments and their interpretations in nonlinear optics 
provide a tool he had not.

To avoid a complicated mathematical development, next subsection shows properties which have been 
demonstrated, but appear more clearly describing experiments.
\subsection{The filaments of light}
In media whose index of refraction increases quickly with the amplitude of an electric field, up to a saturation, a 
powerful laser beam splits into filaments.
These filaments of light which are (2+1)D solitons \footnote{The first index indicates the dimension of the limited 
extension of the wave (here 2 for $x, y$), the second the dimension of its propagation (along $z$).}, have been 
extensively studied experimentally and theoretically \cite{ Chiao,Marburger,Stegeman2,Brodeur,Feit,Zharov}. The 
required nonlinearity is found in many media, in particular Kerr, photorefractive, or plasma; here, we will consider 
the propagation of a monochromatic wave in perfect, homogenous, isotropic, lossless media.

Set $\mbox{\boldmath $E$}(x, y, z, t)$ the electric field, assumed polarised along $Ox$, of the filament propagating 
along the $Oz$ axis, and $E(x, y, z, t)$ the non-zero component of this field. 

\medskip
The electric field $\mbox{\boldmath $E$}(x, y, z, t)$ is an exact solution of Maxwell's equations in which the 
relative permittivity $\epsilon$ is a function of $|\mbox{\boldmath $E$}(x, y, z, t)|$, and the relative permeability 
$\mu$ is 1. Neglecting $|\mbox{\boldmath $\nabla . E$}(x, y, z, t)|$, Helmholtz propagation equation writes
\begin{equation}
\mbox{\boldmath $\Delta E$}(x, y, z, t)=\frac{\mu\epsilon(|E(x, y, z, t)|)}{c^2}{\partial^2 {\bf E}(x, y, 
z, t)\over\partial t^2}.\label{propa}
\end{equation}
Set \cite{Chiao}
\begin{equation}
\mbox{\boldmath $E$}(x, y, z, t)= \mbox{\boldmath $E$}_t(x,y)\cos(kz-\omega t). \label{propag}
\end{equation}
$ \mbox{\boldmath $E$}_t(x,y)$ represents the radial variation of the field and may be written $ \mbox{\boldmath 
$E$}_t(r)$ while $\cos(kz-\omega t)$ is the propagation term; the fields remain unchanged by an increase of the 
coordinate $z$ by a period $\Lambda$. Assuming that the beam is cylindrical, and that there is no absorption, the 
time-reversal invariance shows that $ \mbox{\boldmath $E$}_t(x,y)$ is real, so that the wavefronts are planes 
perpendicular to $Oz$.

Almost all energy of the filament propagates in a cylinder of axis $Oz$ named the core; outside, the nonlinearity is 
negligible and the field is evanescent. The flux of energy in the filament has a well defined \textquotedblleft 
critical" value.

In photorefractive media, the losses of energy are low, so that the filaments are long, and it appears that they are 
very stable; perturbations may modify the evanescent field with only a small change of the critical energy. While 
a symmetrical perturbation leaves the filament straight, an unsymmetrical perturbation curves it, its sections 
perpendicular to the direction of propagation becoming nearly elliptical \cite{Petter}; the origin of a perturbation 
may be the evanescent field of an other filament \cite{Shih} then the filaments curve up to a spiralling 
\cite{Poladian,Belic}.

This behaviour may be interpreted as a refraction of the filament by the polarisation of the medium provided by an 
external field; in a first approximation, the filament is attracted by the regions where the field, thus the index of 
refraction are large. More precisely, the interaction of two filaments depends on the density of energy provided 
by the superposition of the fields, that is on the interference of the fields: a coherence of the fields provides a 
particularly strong interaction which may be repulsive.

A very powerful laser pulse is split into filaments in many media. These filaments start by an accretion of field and 
energy at an initial local maximum of the fluctuating field. During their life the real filaments lose much energy by 
various processes, and the length of their lives shows that they keep their eigen energy by an absorption of an 
available field remaining from the laser pulse or from the death of other filaments: the stability of the filament 
includes a recovering of the critical energy eating available fields at its frequency. As the speed of propagation of 
the filament is not much lower than the speed of light, the propagation of a perturbation is only in a cone having 
its vertex at the perturbation.

If a filament crosses a hole of a screen, and loses a part of its evanescent wave, it may recover it from the zero 
point field. If there is a second hole in the screen, a part of the evanescent field crosses this second hole; both 
parts of the evanescent field interfere into Young's fringes whose maximums attract the filament: the filament 
interferes with itself.

\subsection{Perturbation of a filament by a magnetic nonlinearity.}

Assume a perturbation such that the index of refraction $n(x,y) = \sqrt{\epsilon(x,y)\mu(x,y)}$ increases not only 
with the electric field, but with the modulus $|\mbox{\boldmath $\nabla \wedge E$}(x, y, z, t)|$, that is with the time 
derivative of the modulus of the magnetic field (or with the modulus of the amplitude of magnetic field for a fixed 
frequency). The variation of the permeability may be written, using cylindrical coordinates $(r,\theta,z)$ : 
\begin{equation}
\delta\mu_0=f_0(({\rm curl} \mbox{\boldmath $E$}(r,\theta,z))^2)=f_0\Bigl\{\Bigl (\frac{\partial \mbox{\boldmath 
$E$}}{\partial z}\Bigr)^2+
\frac{1}{r^2}\Bigl(\frac{\partial(r E_\theta)}{\partial r}+\frac{\partial E_r}{\partial 
\theta}\Bigr)^2\Bigr\}\label{mu1}
\end{equation}
Respecting the symmetry, this perturbation does not curve, or modify much, the filament.

\medskip
Consider now an {\it other problem}, in the same homogenous, isotropic medium; to set it, we use curved 
cylindrical coordinates represented in figure 3, $R$ being a constant parameter; with these coordinates, the 
components of the curl are:

\begin{figure}
\begin{center}
\includegraphics[height=6 cm]{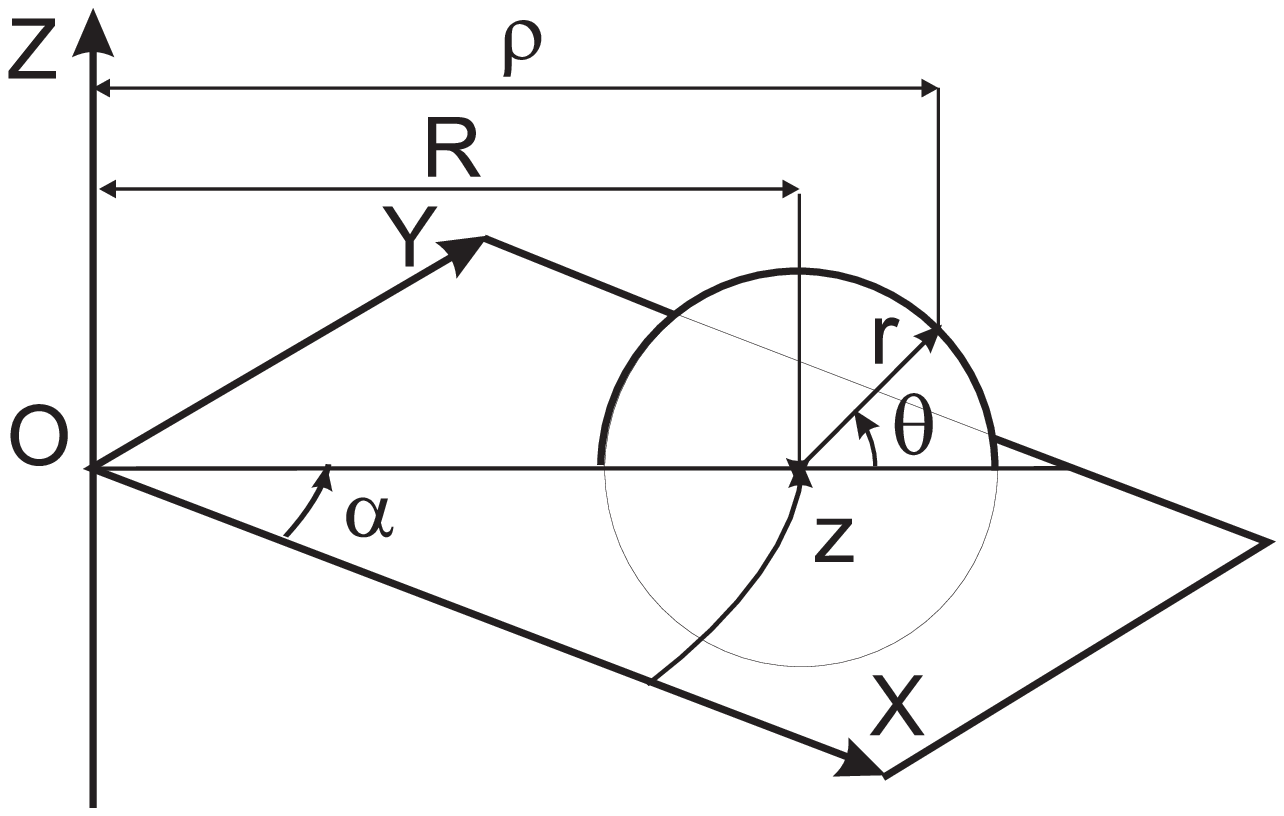}
\end{center}
\caption{Curved cylindrical coordinates}
\end{figure}

\begin{equation}
({\rm curl} \mbox{\boldmath 
$E$})_r=\frac{1}{r(R+r\cos\theta)}\Bigl\{\frac{\partial[(R+r\cos\theta)E_\alpha]}{\partial\theta}- \frac{r\partial 
E_\theta}{\partial \alpha}\Bigl\}
\end{equation}
\begin{equation}
({\rm curl} \mbox{\boldmath $E$})_\theta=\frac{1}{R+r\cos\theta}\Bigl\{\frac{\partial E_r}{\partial \alpha}-
\frac{\partial [(R+r\cos\theta)E_\alpha]}{\partial r}\Bigr\}
\end{equation}
\begin{equation}
({\rm curl} \mbox{\boldmath $E$})_\alpha=\frac{1}{r}\Bigl\{\frac{\partial (r E_\theta)}{\partial r}-\frac{\partial 
E_r}{\partial \theta}\Bigr\}
\end{equation}
Suppose that a daemon sets electric and magnetic fields equals to the value they had in the previous problem for 
equal values of the coordinates having the same name. Taking into account that $E_\alpha$ is null, the 
perturbation induces a variation of the permeability
\begin{equation}
\delta\mu=f(({\rm curl} \mbox{\boldmath $E$}(r, \theta,\alpha))^2)=f\Bigl\{
\frac{1}{(R+r\cos\theta)^2}\Bigl (\frac{\partial \mbox{\boldmath $E$}}{\partial \alpha}\Bigr)^2+
\frac{1}{r^2}\Bigl(\frac{\partial(r E_\theta)}{\partial r}-\frac{\partial E_r}{\partial 
\theta}\Bigr)^2\Bigr\}\label{mu2}
\end{equation}

Equations \ref{mu1} and \ref{mu2} differ by a term, depending on $R/\rho$ with $\rho=R+r\cos\theta$; this term 
produces a decrease of the index of refraction according to $\rho$. Although this variation of the index of 
refraction is not a linear function of $\rho$, the stability of the filament implies that, for a propagation of the wave 
surface corresponding to an increase $\Delta\alpha$ of $\alpha$, the wave surface is turned by an angle 
$\Delta\beta=\gamma\Delta\alpha$ which defines a curvature $\gamma/R$. 

If $\gamma$ equals one, the daemon is useless, a toroidal solution is found. Is it stable? To ease the discussion, 
replace, in equation \ref{mu2}, $R$ by the inverse of the curvature $C=1/R$, so that $\delta\mu$ becomes a 
function of $r, \theta, \alpha$ depending on the parameter $C$. $\Delta\beta$ depends on the parameter $C$ and, 
setting $z = \alpha R$, on $\Delta z= \Delta\alpha/C$.

Study the variation of $\Delta\beta(\Delta z, C)$ for a constant value of $\Delta z$. Choose function $f$ so that it 
increases fast for low values of $C$, then saturates; the variations of the index of refraction, and of $\Delta\beta$ 
are similar; thus, whichever its value for $C=0$, $\Delta\beta(\Delta z, C)$ becomes larger than $\alpha$ for a small 
value of $C$, then $\gamma$ decreases; thus, as $\Delta\beta$ reaches the value $\Delta\alpha$, the derivative 
${\rm d}\Delta\beta/{\rm d}\Delta\alpha$ is lower than one; for the corresponding value $C_0=1/R_0$ of $C$ we 
have a locally stable solution.

The value $R_0$ of the radius of curvature depends only on the properties of the medium and of the wavelength, 
so that the kernel of the filament closes into a torus; the phases of the fields must be the same at the surface of 
junction; the phase may be adjusted, changing the frequency of the wave or the properties of the medium, so that 
$2\pi R_0$ becomes an integer multiple of $\Lambda$; thus, several toroidal solutions of the nonlinear Maxwell's 
equations are found. The demonstration requires that the evanescent field may be neglected for $\rho$ small. 

The flux of energy in the filament being close to the critical value, and the length of filament transformed into a 
torus depending on the assumed properties of the medium, the energy of the soliton is quantified. The (2+1)D 
soliton is transformed into a (3+0)D soliton whose core occupies a limited region of the space, static, or, changing 
the Galilean frame of reference, having any speed, on the contrary of light bullets (stable segments of filaments, 
that is (3+1)D solitons) which move fast.

If the torus moves in relation to a screen and crosses a Young hole, its evanescent field is cut up and makes 
interferences; these interferences modify the index of refraction, so that the trajectory of the torus follows the 
interferences: a large population of solitons hitting the screen draw Young's interferences.

\medskip

The existence of electromagnetic (3+0)D solitons has been shown, apparently for the first time, in the particular 
case where the variation of the index of refraction produced by the magnetic field is low; it may be a starting point 
for numerical computations of more general solitons; numerical computations seem necessary to answer many 
questions such as:

- what happens increasing the magnetic nonlinear contribution to the index of refraction: in particular, can the 
torus become next to a sphere ?

- can the soliton have an electric or magnetic charge introduced by a non zero divergence of the fields ?

\section{Tentative setting of a classical theory including the important quantum results}
A classical interpretation of the properties of the particles must solve many problems; among those is the 
necessity to represent a large number of particles, then a new theory must be connectable to the old ones.
\subsection{Is matter made of electromagnetic (3+0)D solitons ?}
The wave constituting the soliton may be, somewhat artificially, split into a linear, evanescent part and the 
nonlinear part. These parts may be identified with the $\psi$ and $u$ waves of the \textquotedblleft double 
solution" if the vacuum has convenient non-linear properties. Usual optics in the vacuum shows that vacuum is 
strictly linear. De Broglie suggested the introduction of nonlinearities; more precisely, Schwinger 
\cite{Schwinger} supposed that the vacuum is nonlinear; high energy $\gamma$ rays propagating in a magnetic 
field may create electron positron pairs \cite{Ritus}; this implies that Maxwell's equations become non linear, so 
that supposing that the vacuum has nonlinearities is not absurd. Thus the particles may be (3+0)D solitons similar 
to those introduced in the previous section. The properties of the ring model of the electron \cite{Allen} bound 
to its symmetries could apply to this soliton.

The demonstration of the existence of (3+0)D solitons is founded on a perturbation calculation in which the 
starting point is a filament. Is the filament a mathematical tool only, or does it has a physical interpretation? It may 
be a string whose mechanical properties depend not on extra dimensions, but on nonlinearities. The use of 
nonlinearities, seems a powerful tool because the theory of the curves is simple, and it allows to introduce a 
torsion, just as the curvature was introduced, and complicated functions of curvature and torsion. 

With this identification, the theory of strings gets a physical foundation, while de Broglie's double solution gets a 
good mathematical interpretation.

The neutrinos appear as pieces of filament moving nearly at the speed of the light, \textquotedblleft light bullet" 
\cite{Enns,Monot,Braun}, whose energy depending on the their length is not quantified. As they move fast, a 
concatenation of two bullets , or an integration of a bullet into a closed, static string appears difficult; the difficult 
interaction of the neutrinos with matter is justified.

Fred Hoyle's continuous creation of matter could be an appearance of nonlinearities transforming high frequency 
zero point waves into solitons and solving simultaneously the UV divergence.

As Maxwell's equations in the vacuum are invariant by a commutation of the electric and magnetic fields, we can 
build by this commutation a probably different family of solitons. However, the electric and magnetic fields do not 
play exactly the same role, so that this hypothesis is probably useless.
\subsection{Inserting the quantum calculation of energies into the classical theory.}
The practical start of quantum mechanics, is the Schr\"odinger's equation, and the physicists had to solve partial 
differential equations. The mathematicians proposed powerful methods to find solutions: the first was the use of 
the commutation rules of linear differential operators (equivalent to a matrix algebra), the best is building a Lie 
algebra from these operators. Lie algebra work so well that they become the foundation of the theories in the 
fields of elementary particles and in spectroscopy.

Quantum mechanics found very remarkable results such as a frequent dependence of energy on a function 
$j(j+1)$ of a quantum number $j$. Lie algebra show that this property is simply a consequence of the symmetry of 
the space. Finding the energy levels of a system is now setting a Lie algebra partly from assumed symmetries, 
partly arbitrarily; then an \textquotedblleft hamiltonien operator" is a function of the generators of the algebra. 
The algebra and the hamiltonian are chosen to obtain the observed eigenvalues, and it appears clearly that, taking 
into account the choice of a \textquotedblleft good algebra", the really introduced number of parameters is of the 
order of the number of observations; it is accepted to include other mathematical methods in quantum mechanics, 
as they appear sometimes more fruitful, for instance using Padé's approximants 
\cite{Langhoff,Goscinski,Goodson}.

Thus, our powerful spectroscopy is only a very good method of interpolation and extrapolation in the space of 
the quantum numbers. It may be considered as a phenomenological classical way to represent the minimums of 
potential of a complicated, unsolved problem of classical electromechanics.

The other tools of quantum mechanics are often founded on classical computations, they use a lot of good 
recipes, but none of the useful tools needs the Copenhagen principles.

\section{Conclusion}
Quantum theory was the source of major improvements in Optics, spectroscopy, and , more generally, physics. It 
was widely developed at the beginning of the 20th century, as a substitute of classical physics which appeared 
limited for various reasons: 

i) the electromagnetism studied isolated simple systems which ignored the zero point field; 

ii) linear theories led to the important discoveries in the electromagnetism at the end of the 19th century, so 
that Poincar\'e' theories about the nonlinear systems appeared too difficult in comparison with their 
forseeable applications, so that they were not used; 

iii) the discovery of some elementary particles was recent, so that it was a temptation to find particles everywhere.

All proved progresses of quantum mechanics were initiated using Schr\"odinger's equations, Heisenberg's and 
Dirac's matrices, or other methods which did not result from precise and coherent postulates. Copenhagen 
postulates tried to introduce a coherence into the theory, but they led to paradoxes which showed their 
incoherence. Up to now, it was impossible to verify any specific prediction deduced from Copenhagen 
postulates; worse, the darkness of the postulates led many physicists to ignore the importance of the sub-
quantum exchanges of energy between non stationary systems, so that they tried to discourage Townes from 
discovering the laser\cite{Townes}, or prevented the astrophysicists from imagining that a simple light-matter 
interaction \cite{Mor1} may be confused with a Doppler effect.

The progresses of mathematical methods and experimental optics allow to overtake the problems which led to 
replace a powerless classical physics by a theory belatedly founded on some magic. Replacing postulates by 
demonstrations is not easy; while this problems is solved in optics by an elimination of the photon as a particle, 
we have only a canvas of the elimination of the wave-particle duality through an identification of de Broglie's 
double solution with a three dimensional static soliton. A fragile way opens to a correlation between the double 
solution and the theory of strings, but it remains much work before a strong settlement.

Fred Hoyle's continuous creation of matter could be a transformation of high frequency zero point waves into 
solitons, solving simultaneously the UV divergence and opening a large topic to astrophysics.


\end{document}